\documentclass[]{spie}  

\usepackage{amsmath,amsfonts,amssymb}
\usepackage{graphicx}
\usepackage[colorlinks=true, allcolors=blue]{hyperref}

\title{VERTECS: A COTS-based payload interface board to enable next generation astronomical imaging payloads}

\author[a]{Ezra Fielding}
\author[a]{Victor H. Schulz}
\author[a]{Keenan A. A. Chatar}
\author[a]{Kei Sano}
\author[a]{Akitoshi Hanazawa}
\affil[a]{Kyushu Institute of Technology, 1-1 Sensui-cho, Tobata-ku, Kitakyushu, Japan}

\authorinfo{Further author information: (Send correspondence to E.F.)\\E.F.: E-mail: fielding.ezra455@mail.kyutech.jp}

\begin{document} 
\maketitle

\begin{abstract}
Due to advances in observation and imaging technologies, modern astronomical satellites generate large volumes of data. This necessitates efficient onboard data processing and high-speed data downlink. Reflecting this trend is the Visible Extragalactic background RadiaTion Exploration by CubeSat (VERTECS) 6U Astronomical Nanosatellite. Designed for the observation of Extragalactic Background Light (EBL), this mission is expected to generate a substantial amount of image data, particularly within the confines of CubeSat capabilities. This paper introduces the VERTECS Camera Control Board (CCB), an open-source payload interface board leveraging Commercial Off-The-Shelf (COTS) components, with a Raspberry Pi Compute Module 4 at its core. The VERTECS CCB hardware and software have been designed from the ground up to serve as the sole interface between the VERTECS bus system and astronomical imaging payload, while providing compute capability not usually seen in nanosatellites of this class. Responsible for mission data processing, it will facilitate high-speed data transfer from the imaging payload via gigabit Ethernet, while also providing a high-bitrate serial connection to the payload X-band transmitter for mission data downlink. Additional interfaces for secondary payloads are provided via USB-C and standard 15-pin camera connectors. The Raspberry Pi embedded within the VERTECS CCB operates on a standard Linux distribution, streamlining the software development process. Beyond addressing the current mission's payload control and data handling requirements, the CCB sets the stage for future missions with heightened data demands. Furthermore, it supports the adoption of machine learning and other compute-intensive applications in orbit. This paper delves into the development of the VERTECS CCB, offering insights into the design and validation of this next-generation payload interface, to ensure that it can survive the rigors of space flight.  
\end{abstract}


\section{INTRODUCTION}
\label{sec:intro}
The Visible Extragalactic background RadiaTion Exploration by CubeSat (VERTECS) mission \cite{sano2023vertecs} aims to reveal star formation history by observing Extragalactic Background Light (EBL) using a 6U ($10$cm$\times{20}$cm$\times{30}$cm) nanosatellite based on the open-source BIRDS Standard CubeSat Bus\cite{kim2021birds} and the KITSUNE 6U CubeSat\cite{orger2022kitsune}. A CubeSat is a class of nanosatellite developed by a number of universities and organizations around the concept of providing small, low construction cost, low launch cost space experiment platforms\cite{heidt2000cubesat}.

CubeSats have emerged as a significant advancement in enabling space-based telescopes\cite{chatar2023downlink}. There have been a number of recent and upcoming nanosatellite projects, which use the same 6U form factor as VERTECS, undertaking missions of an astronomical nature. ASTERIA featured enabling technologies for high-precision photometry in a 6U CubeSat form factor, demonstrating stable milliKelvin-level focal plane thermal control and arcsecond-level line-of-sight pointing stability\cite{knapp2018asteria}. The CUTE CubeSat mission was designed to observe near-ultraviolet stellar brightness distribution to conduct a survey of the extended atmospheres of nearby close-in planets\cite{france2023cute}. Meanwhile, ZODIAC is a proposed upcoming 6U CubeSat designed to detect the near-infrared wavelength to study the structure and dynamics of interplanetary dust cloud though spectral imaging of the zodiacal light\cite{zemcov2022zodiac}. All of these missions appear to make use of proprietary satellite bus systems, usually procured from an external supplier or developed in-house.

The primary objective of the VERTECS 6U Astronomical Nanosatellite is to conduct spectral observation in the visible wavelength ($0.4\mu$m--$0.8\mu$m) to reveal the origin of the EBL. VERTECS, like other astronomical nanosatellites \cite{knapp2018asteria, france2023cute, zemcov2022zodiac}, will generate a large volume of data due to the inclusion of a modern high-resolution imaging sensor. Efficient onboard data processing and high-speed data downlink is required to ensure the successful completion of the mission within the required time-frame. The limitations imposed by the CubeSat form factor have a direct impact on the onboard storage and computational power available for onboard data processing and analysis\cite{chatar2024onboard}. These are factors which need to be considered when designing the interface for high data volume producing payloads, such as astronomical imaging payloads.

VERTECS will derive the majority of its space-flight heritage from the KITSUNE 6U CubeSat. KITSUNE featured multiple missions ranging from earth observation to technology demonstrations\cite{orger2022kitsune}. Most notably, KITSUNE included a 5m class resolution imaging payload, featuring a 31.4 megapixel CMOS sensor, primarily used to capture man-made patterns on the ground by groups of people. To control this imaging payload, KITSUNE included a Camera Control Board (CCB), designed around the Raspberry Pi Compute Module 3 (CM3), which allowed for processing, storage and downlinking of images via the included C-band transmitter\cite{azami2022earthobs}. This CCB served as the main payload interface between the KITSUNE bus and the imaging payload. VERTECS will build off of and adapt the KITSUNE platform to enable its astronomical mission.

This paper presents the completed open-source VERTECS Camera Control Board (CCB), first conceptualized in Ref.~\citenum{chatar2023downlink}. The VERTECS CCB will serve as the sole interface between the VERTECS bus and camera payload. It builds off of the heritage of the KITSUNE CCB\cite{azami2022earthobs}, while upgrading its capabilities to meet the high demands of the VERTECS mission\cite{sano2023vertecs}. Designed with future re-usability in mind, Commercial Off-The-Shelf (COTS) components and open-source Electronic Computer-aided Design (ECAD) software were used during the development process. Section~\ref{sec:satdesign} will present an overview of the design of the VERTECS 6U nanosatellite, while Section~\ref{sec:sysdes} describes how the VERTECS CCB acts as the bridge between the payload and bus systems, as well as presenting the hardware and software choices made during the design phase. The methods used to validate the design of the VERTECS CCB will be presented in Section~\ref{sec:methods}, followed by a discussion of the results of these methods in Section~\ref{sec:results}. A conclusion will be drawn in Section~\ref{sec:conclusion}. The VERTECS CCB design files, schematics and parts list will be available on GitHub as an open-source project\footnote{VERTECS CCB GitHub repository: \url{https://github.com/ezrafielding/vertecs-ccb}}.

\section{SATELLITE DESIGN}
\label{sec:satdesign}
\begin{figure} [ht]
   \begin{center}
   \begin{tabular}{c}
   \includegraphics[height=5cm]{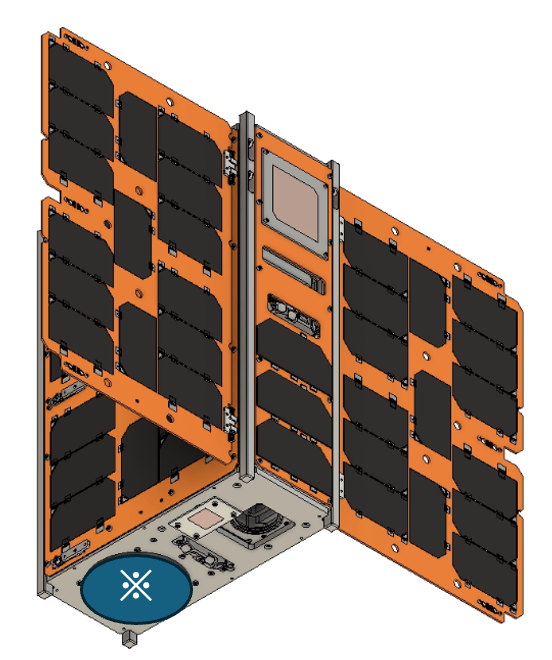}
   \end{tabular}
   \end{center}
   \caption[VERTECS] 
   { \label{fig:vertecs} 
Depiction of the VERTECS 6U Astronomical Nanosatellite with solar panels deployed.}
\end{figure}

The VERTECS nanosatellite, depicted in Fig.~\ref{fig:vertecs}, is designed to conform to the 6U CubeSat form factor ($10$cm $\times{20}$cm$\times{30}$cm). As mentioned in Section~\ref{sec:intro}, VERTECS derives a large majority of its space-flight heritage from the KITSUNE 6U CubeSat\cite{orger2022kitsune}. The main structure and satellite bus system follow closely to those implemented in KITSUNE, with minor revisions and updates where required. A new camera payload and telescope will be developed to enable the scientific objectives of the VERTECS mission.

\subsection{VERTECS Bus}
The latest version of the BIRDS Standard CubeSat Bus\cite{kim2021birds}, developed at Kyushu Institute of Technology, will be used as the VERTECS bus system, following on from the legacy of KITSUNE. The VERTECS bus features a revised Onboard Computer (OBC) board, designed to be more tolerant to faults during initial operation, along with a new back-plane board to facilitate the addition of new communications and attitude determination control system (ADCS) hardware to enable the VERTECS mission. VERTECS will include an X-band transmitter for high-speed mission data downlink, while an S-band transmitter and receiver will be used for house-keeping data downlink and command uplink. To achieve the high-precision pointing required to enable the mission, VERTECS will include Blue Canyon Technology's XACT ADCS system.

\subsection{Payload}
The main camera payload is comprised of a custom made telescope and camera module, featuring a Sony IMX533 monochrome imaging sensor. Communication for sending commands to and receiving data from the camera module is done via a gigabit Ethernet connection. Power is provided to the module via the same surface mount connector as the gigabit Ethernet connection. Given the 3003 pixel by 3008 pixel size of the IMX533 sensor, at 16 bits of data per pixel, each frame captured by VERTECS will produce 17.23MB of data. To achieve full mission success within 1 year, 40 frames must be captured per day, resulting in a data generation rate of about 689MB per day, a large amount of data by CubeSat standards.

VERTECS will also include a Raspberry Pi Camera Module 3 as an earth-facing secondary camera for engineering and earth observation purposes. The Raspberry Pi Camera Module 3 includes a 11.9 megapixel Sony IMX708 sensor.

No previous project utilizing the BIRDS CubeSat Bus has dealt with the amount of data expected from VERTECS, nor has any previous project had to deal with such high-speed communications with the payload and communications system for mission data downlink. Therefore, a new payload control interface is required, capable of high-speed data transfer and processing, to enable the VERTECS bus to complete the VERTECS mission.

\section{VERTECS CCB DESIGN}
\label{sec:sysdes}
Given the data-intensive nature and strict requirements of the VERTECS mission, the VERTECS CCB should fulfill a specific set of system requirements to ensure that it is capable of managing the various payload interfaces, as well as the large volume of data that are expected to be generated. This is expressed in the following system requirements:
\begin{enumerate}
    \item CCB shall control the mission payload camera, including status check, power line check, exposure, and data collection.
    \item CCB shall control the secondary camera, including status check, power line check, exposure, and data collection.
    \item CCB shall control the X-band transmitter, including status check, power line check, and all data transmission.
    \item CCB shall send data to the X-band transmitter for downlink at a data-rate greater than or equal to 5 Mbps.
    \item CCB shall have the data storage capacity for three-day observation data.
\end{enumerate}
These requirements are considered in conjunction with the general requirements imposed on a CubeSat which cover the survivability of components in the rocket launch and space environments. Requirements 1, 2 and 5 cover the control and data management of the primary and secondary payload cameras. Meanwhile, requirements 3 and 4 deal with the control of the X-band transmitter for mission data downlink, while setting a minimum attainable downlink rate to ensure that the main mission is completed within a reasonable time in orbit. 

\begin{figure} [ht]
   \begin{center}
   \begin{tabular}{c}
   \includegraphics[height=8cm]{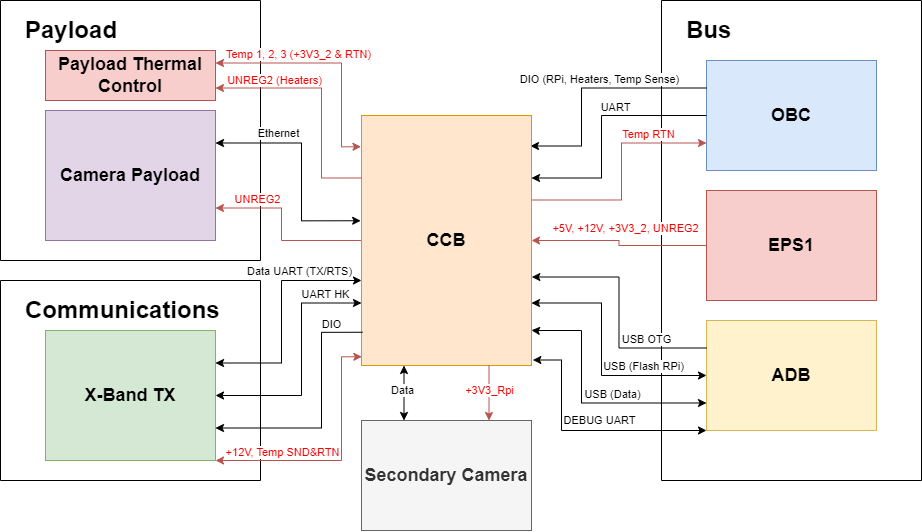}
   \end{tabular}
   \end{center}
   \caption[System Overview] 
   { \label{fig:systemlevel} 
System-level overview diagram depicting the connections between the VERTECS bus, payload and CCB.}
\end{figure}

The VERTECS CCB is designed to serve as the sole bridge between the VERTECS bus and payload, responsible for providing power and data communications for all payload related components (including those related to  mission data downlink). Fig.~\ref{fig:systemlevel} illustrates the connections between the CCB and the various VERTECS subsystems.  

\subsection{Hardware}
The VERTECS CCB hardware was designed around the use of Commercial Off-The-Shelf (COTS) components using KiCAD, an open-source Electrical Computer-aided Design (ECAD) software. Using COTS components reduces the overall component cost and, coupled with the use of open-source ECAD software, promotes future re-usability and development. The CCB is designed to sit on the payload side of the VERTECS nanosatellite, mounted to the payload assembly and connected to the VERTECS bus back-plane board via a 30-pin harness. The board measures $90$mm$\times{74}$mm$\times{1.6}$mm.

The Raspberry Pi Compute Module 4 (CM4) with 32GB of eMMC flash storage, 8GB of RAM and wireless capabilities was selected as the main compute unit (MCU) for the VERTECS CCB. The CM4 was chosen over the Raspberry Pi Compute Module 3 (CM3) found in the KITSUE CCB due to the CM4 being more readily available, an increase in available compute power, and the inclusion of integrated gigabit Ethernet capabilities. The CM4 provides more interface options, which are useful for future expansion. The inclusion of the CM4 also supports the future adoption of machine learning and other compute-intensive applications in orbit. Ref.~\citenum{chatar2023downlink} and Ref.~\citenum{chatar2024onboard} present and explore machine learning algorithms designed for and run on CM4 class compute units on a CubeSat in orbit. The VERTECS CCB, with its included CM4, will allow for and further encourage innovation along these lines by providing a flight-proven platform for future developments to make use of. The included wireless capabilities of the CM4 will only be used during software development and will be hardware disabled before flight. The CM4's 32GB of eMMC storage will fulfill requirement 5 which sets the amount of mission data storage required.

Power and a gigabit Ethernet connection, via the built in networking capability of the CM4, are provided to the main camera payload through a 20-pin surface connector. The VERTECS CCB will be mounted to and mated with the camera payload module via four M3 screw holes and the 20-pin surface connector, respectively.

To enable high-speed data transfer to the X-band transmitter, the CCB includes an FTDI FT232H chip connected to the CM4 via internal USB. This allows the CCB to communicate with the X-band transmitter via RS232 UART at a rate of 12 Mbaud. An AD7490 analogue-to-digital (ADC) converter was added to the CCB to collect accurate temperature readings from the X-band transmitter's integrated temperature sensor. A 4-pin connector will facilitate the data connection to the X-band transmitter, while a 7-pin connector will be responsible for providing power.

\begin{figure} [ht]
   \begin{center}
   \begin{tabular}{c}
   \includegraphics[height=8cm]{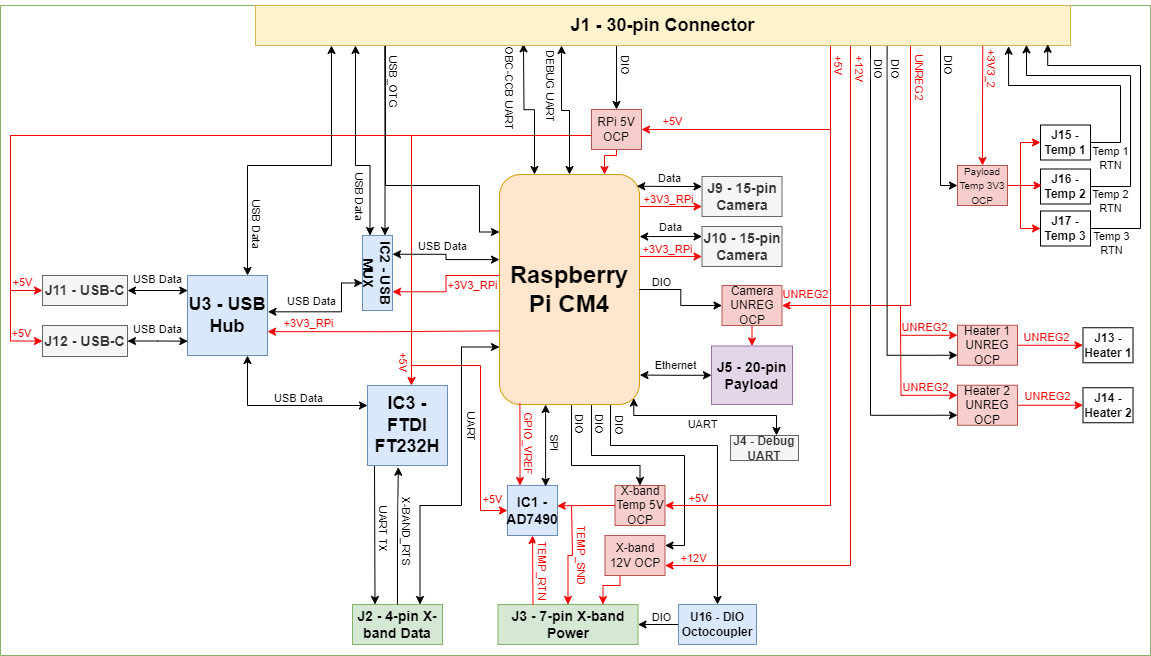}
   \end{tabular}
   \end{center}
   \caption[Subsystem Overview] 
   { \label{fig:subsystem} 
Subsystem-level overview diagram showing connections of components included on the CCB.}
\end{figure}

Two 15-pin CSI camera connectors allow up to two CSI cameras to be connected to the CM4, one of which will be used for the secondary earth-facing camera. Two USB-C ports are included on the CCB to allow for easier development, as well as additional payload capabilities for the satellite. Three 2-pin connections for temperature and two 2-pin connections for heaters are provided on the CCB for control by the main satellite bus system. Over-current protection (OCP) circuits are included for each power line running to the CCB. Payload related OCPs are controlled by the CM4, while the OCPs for the CM4 and payload thermal control system are controlled by the VERTECS bus. Fig.~\ref{fig:subsystem} presents a diagram outlining the subsystem-level connections of the VERTECS CCB.

\subsection{Software}
A number of Linux distributions are available for the Raspberry Pi CM4. Raspberry Pi OS Lite was chosen for this project due to its minimal image size, stable performance, and good compatibility with the CM4's built in hardware. It is envisaged that the majority of software development for the CCB will take place using Python 3, due to its easy to read syntax and good integration with Raspberry Pi functions. In the case where more performance is required, the C programming language will be used. Software development will be done primarily using Visual Studio Code's remote connection feature via SSH over WiFi.

\section{VALIDATION METHODS}
\label{sec:methods}
\begin{figure} [ht]
   \begin{center}
   \begin{tabular}{c}
   \includegraphics[height=6cm]{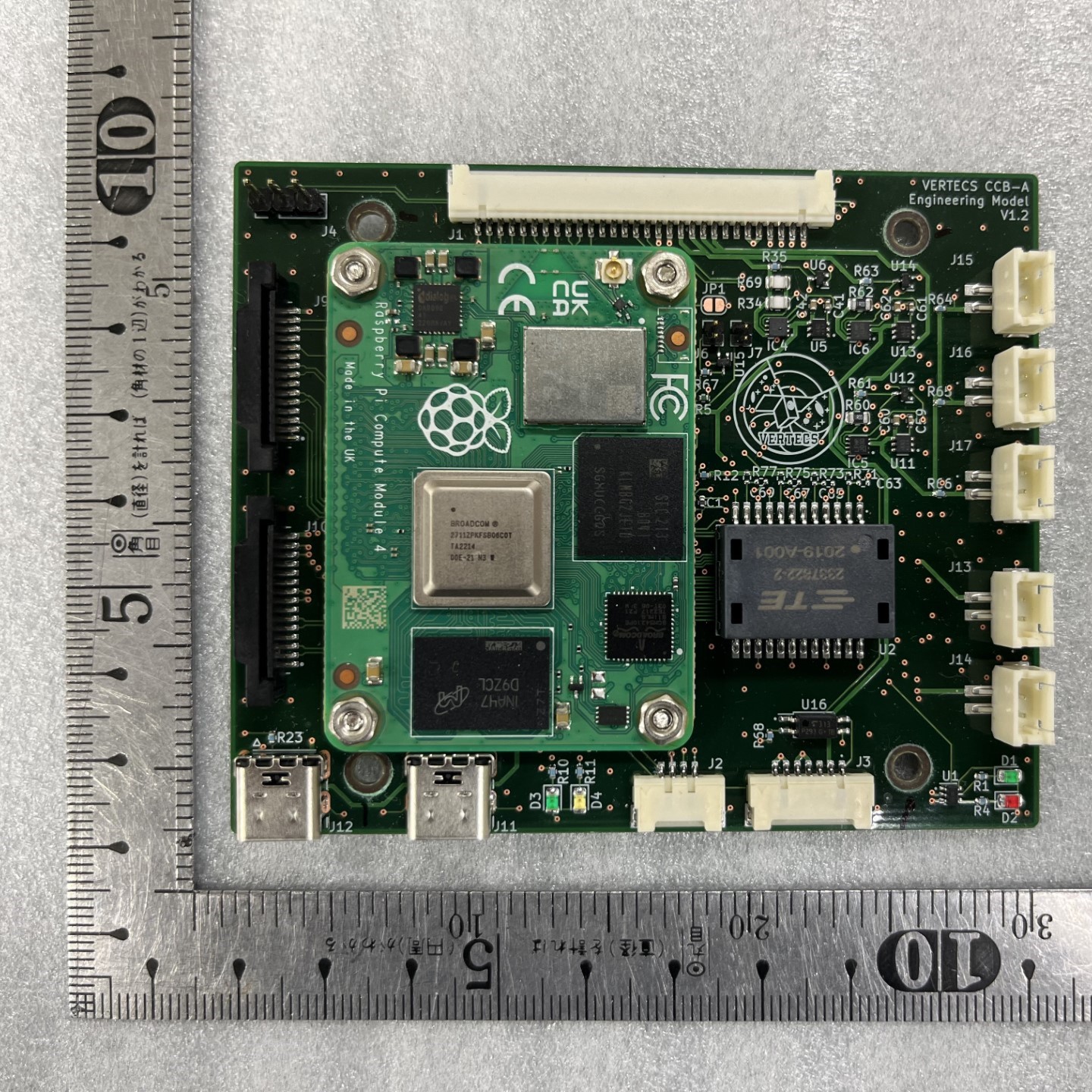}
   \includegraphics[height=6cm]{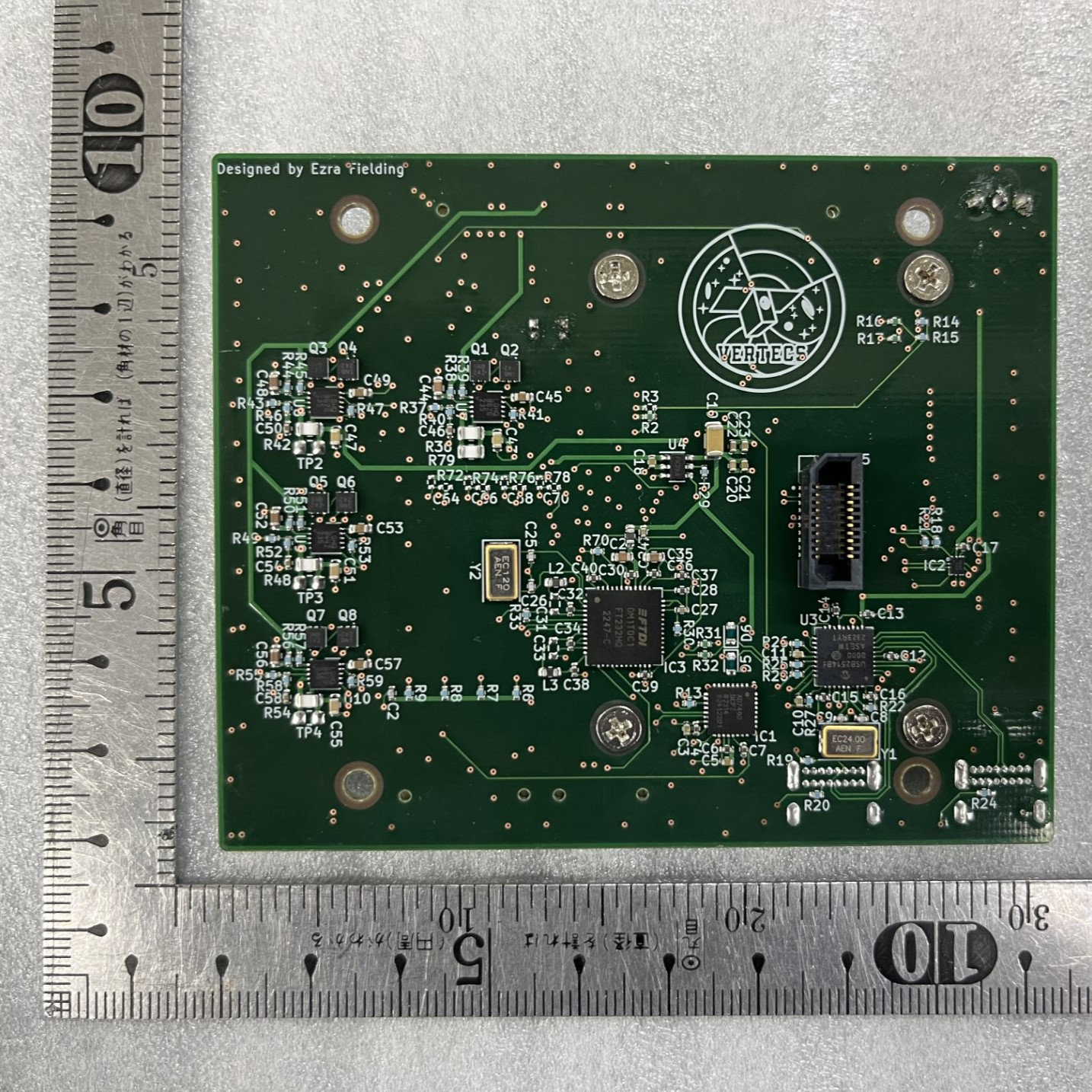}
   \end{tabular}
   \end{center}
   \caption[CCB] 
   { \label{fig:ccbimage} 
Front (left) and back (right) of the VERTECS CCB Engineering Model.}
\end{figure}

\begin{figure} [ht]
   \begin{center}
   \begin{tabular}{c}
   \includegraphics[height=5cm]{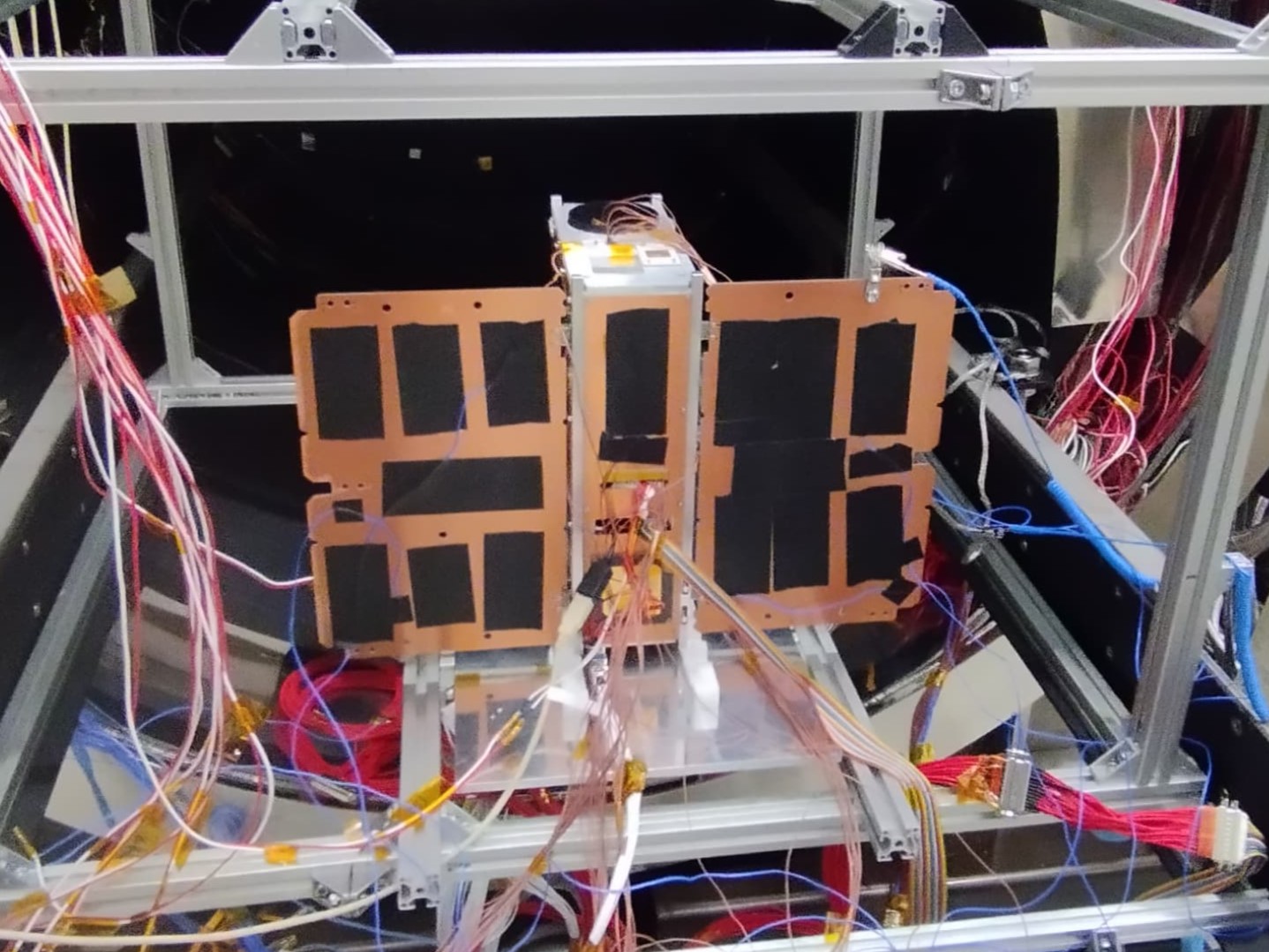}
   \end{tabular}
   \end{center}
   \caption[TVT Configuration] 
   { \label{fig:tvtconfig} 
VERTECS STM prepped for TVT.}
\end{figure}

The following section will discuss the methods used to validate the VERTECS CCB design to ensure that it can survive the rigors of space-flight while meeting the strict requirements of the VERTECS mission. The VERTECS CCB Engineering Model (EM), seen in Fig.~\ref{fig:ccbimage}, will be used to complete the validation tests. The majority of the validation testing will take place in the table-sat configuration, where VERTECS components are connected to each other on a table before integration into the satellite structure. Some validation testing will take place during the Thermal Vacuum Test (TVT) for the VERTECS Structural and Thermal Model (STM), with the satellite components integrated into the STM structure, as seen in Fig.~\ref{fig:tvtconfig}.

The tests performed during the TVT are essential for confirming the survivability of the COTS components in the space environment. TVT will test the components in a vacuum environment, simulating direct sunlight and eclipse conditions. The direct sunlight condition will subject the satellite to an ambient temperature of $30^{\circ}$C, while the eclipse condition will have an ambient temperature of $-20^{\circ}$C.

For clarity, the validation methods have been split into subsections relating to the bus interface, payload interface and communications interface.

\subsection{Bus Interface}
To validate the interface between the VERTECS bus and CCB, the functionality of all 30 lines to the back-plane board will be tested. All power lines must carry the correct voltage, while all data lines must be capable of transferring data at the appropriate rate between the CCB and target bus subsystem. This includes communication with the OBC and the Access Deployment Board (ADB). Tests confirming that the VERTECS bus is capable of providing sufficient power to sustain the CCB must be conducted. Control of the payload thermal control devices (temperature sensors and heaters) by the ADB will also be confirmed.

\subsection{Payload Interface}
The interface between the VERTECS CCB and main camera payload will be validated by confirming that the CCB is able to power the camera unit and able to initiate a gigabit Ethernet connection. The ability to send commands and receive data over the Ethernet connection will also be confirmed.

The CCB will be essential for operating the camera payload during the STM TVT. During the eclipse condition phase, the camera payload will undergo 5 sets of 1 minute and 5 sets of 37 millisecond exposures each at 3 different camera gains. This will be repeated 4 times back-to-back. In the direct sunlight condition phase, the same test will be conducted, however, this time skipping the 37 millisecond exposures. The CCB will be on and controlling the camera payload throughout these tests.

The secondary camera interface will be validated by confirming that images and video can be captured and stored by the CM4.

\subsection{Communications Interface}
Validation of the interface between the VERTECS CCB and X-band transmitter will be done by confirming that the CCB is able to provide power to the transmitter and able to transfer data at the required baud-rate. The ability to supply data to ensure downlink at an effective rate of 5Mbps should also be confirmed. Finally, the ADC included on the CCB should be tested to confirm that the X-band transmitter's integrated temperature sensor can be read.

\section{RESULTS}
\label{sec:results}
\begin{figure} [ht]
   \begin{center}
   \begin{tabular}{c}
   \includegraphics[height=4cm]{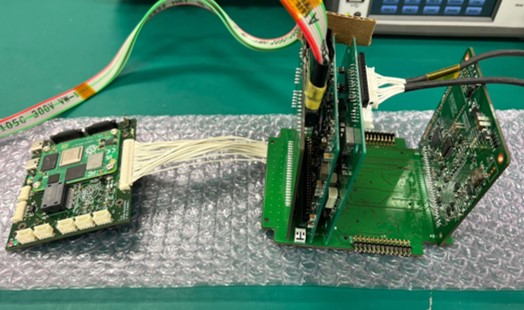}
   \includegraphics[height=4cm]{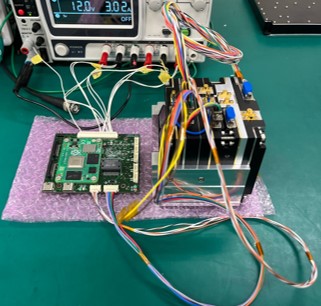}\\
   \includegraphics[height=4cm]{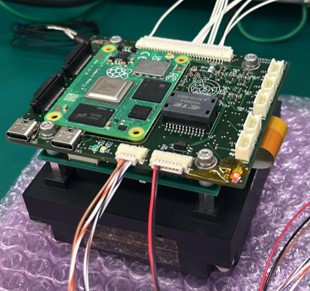}
   \includegraphics[height=4cm]{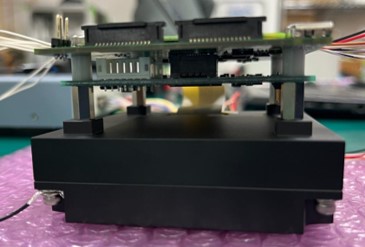}
   \end{tabular}
   \end{center}
   \caption[Table-sat Configuration] 
   { \label{fig:ccbtestsetup} 
Table-sat configurations used for CCB validation tests. Top left shows the CCB connected to the VERTECS bus, top right shows the CCB connected to the X-band transmitter in the communications stack, bottom left shows the CCB attached to the camera module from a top view and bottom right shows the side view.}
\end{figure}

The VERTECS CCB successfully underwent validation tests to ensure that the requirements listed in Section~\ref{sec:sysdes} are met by the current design. The results of these tests will be presented and discussed in the following section. Fig.~\ref{fig:ccbtestsetup} shows the configurations used for testing the VERTECS CCB in the table-sat configurations.

\begin{figure} [ht]
   \begin{center}
   \begin{tabular}{c}
   \includegraphics[height=5cm]{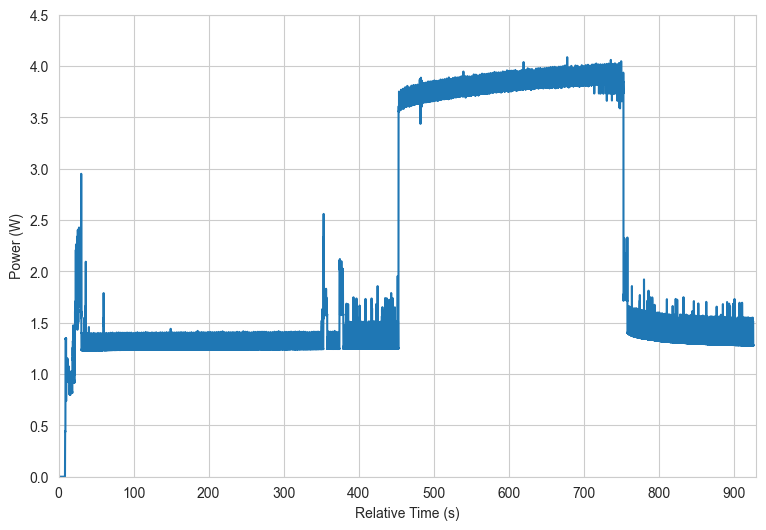}
   \includegraphics[height=5cm]{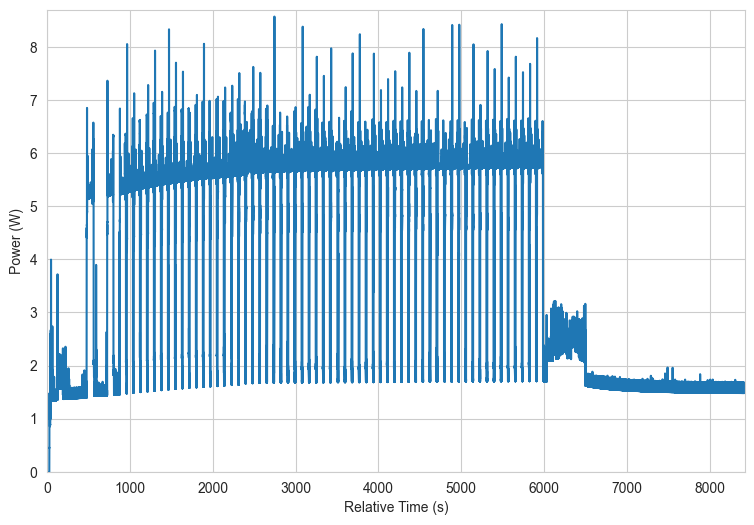}
   \end{tabular}
   \end{center}
   \caption[Power] 
   { \label{fig:power} 
The power-draw characteristics of the CCB in the table-sat configuration (left) and the combined power consumption of the CCB and camera payload during the TVT (right).}
\end{figure}

The graph on the left of Fig.~\ref{fig:power} shows the power-draw characteristics of the CCB when the CM4 is powered on, left to idle and under load in the table-sat configuration. After powering on the CM4, it was left to idle for 5 minutes. This was then followed by 5 minutes of simulated load, which placed all 4 CPU cores at $100\%$ utilization. At idle, the CCB consumes a mean of about 0.25A at 5V (1.25W). Under load, the CCB consumes a mean of about 0.75A at 5V (3.75W). The largest consumer of power on the CCB is the CM4.

The VERTECS CCB successfully underwent TVT, performing nominally in both eclipse and direct sunlight conditions. This indicates that the included COTS components can survive the rigors of space-flight.

\subsection{Bus Interface}
The harness lines connecting the VERTECS CCB to the bus system, including power, data and analogue lines, were checked and validated. All power lines were confirmed to carry the correct voltage and the VERTECS bus was capable of powering the CCB. UART communication between the OBC and CCB was confirmed. Lines related to heater control and temperature sensors were confirmed working and capable of control by the ADB. External access to the CCB via the ADB through UART and USB was also confirmed.

\subsection{Payload Interface}
The VERTECS CCB was able to provide power to the camera payload. Once powered on, the camera payload was able to initiate a gigabit Ethernet connection with the CCB. Commands and data were successfully transferred via this connection and frames captured by the camera payload were stored on the CM4 for later analysis. This was achieved both in the table-sat configuration, as well as during the TVT.

The graph on the right of Fig.~\ref{fig:power}, represents the combined power consumption of the CCB and the camera payload during the TVT test described in Section~\ref{sec:methods}. The graph was calculated by measuring the power consumption of the entire satellite and subtracting the average 1W of power consumed by the VERTECS bus. During the test, the CCB and camera payload consumed a combined mean of about 5.18W.

\begin{figure} [ht]
   \begin{center}
   \begin{tabular}{c}
   \includegraphics[height=5cm]{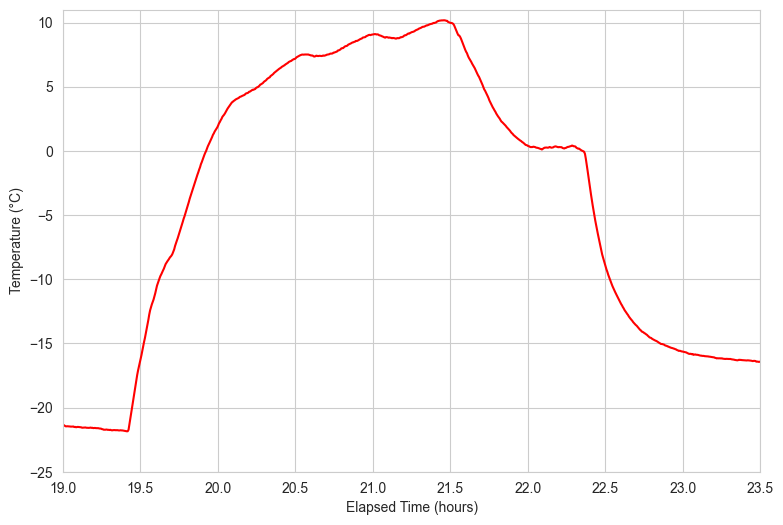}
   \includegraphics[height=5cm]{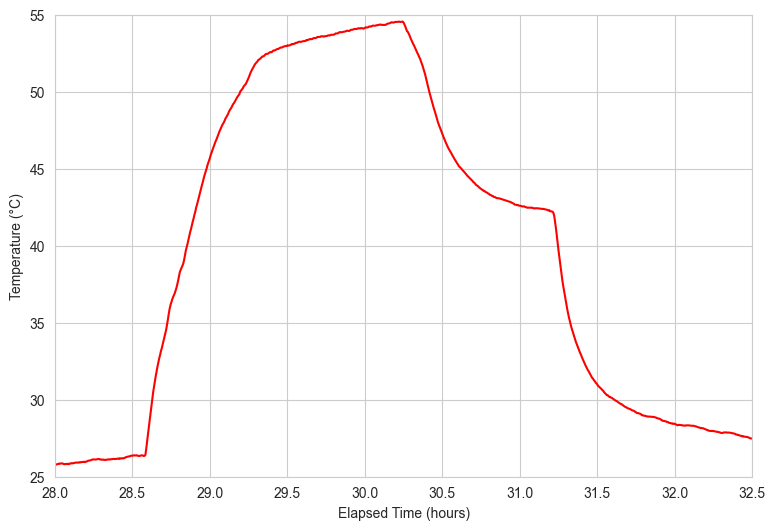}
   \end{tabular}
   \end{center}
   \caption[CCB Temperature] 
   { \label{fig:ccbtemp} 
The temperature of the CCB in the eclipse condition (left) and the direct sunlight condition (right).}
\end{figure}

Fig.~\ref{fig:ccbtemp} shows the temperature of the CCB during the camera payload test in eclipse and direct sunlight conditions during the TVT. The temperature probe for the CCB was placed on the CM4's CPU which is the source of the most heat on the CCB. In the eclipse condition, the maximum temperature of the CCB was about $10.18^{\circ}$C. While in the direct sunlight condition, the maximum temperature of the CCB was about $54.55^{\circ}$C. In both cases, the CCB was turned on once satellite temperatures reached ambient conditions. The test was then run and, once completed, the CCB was left to idle until temperatures stabilized. The CCB was then powered down and the temperature was left to stabilize once again. The temperatures observed are within the CM4's standard operating range.

Still images and videos were successfully captured by the CM4 using the earth-facing Raspberry Pi Camera Module 3. This was achieved both in the table-sat configuration and during the TVT.

\subsection{Communications Interface}
Power for the X-band transmitter was successfully provided via the VERTECS CCB. Once powered on, mission data, house keeping data and command data communication via UART was confirmed. Mission data packets were transferred at a rate which enabled downlink at 10Msps with BPSK modulation, which provides an effective data-rate of 5Mbps during downlink. However, occasional desynchronization of data packets for large files ($>2.4$MB) was observed, suggesting that either further software fine-tuning is required or that the data packet transfer scheme between the CCB and X-band transmitter needs to be reevaluated.

The CM4 was able to use the AD7490 ADC included on the CCB to read the temperature data provided by the X-band transmitter's integrated temperature sensor.

\section{CONCLUSION}
\label{sec:conclusion}
In conclusion, the open-source and COTS-based VERTECS Camera Control Board has been presented and shown to be a device capable of meeting the requirements to enable the VERTECS mission. VERTECS aims to reveal star formation history by observing Extragalactic Background Light (EBL), specifically by conducting spectral observation in the visible wavelength ($0.4\mu$m--$0.8\mu$m). In so doing, VERTECS will produce a large volume of data, about 689MB per day. The VERTECS CCB has been designed to serve as the bridge between the VERTECS bus and payload, building on from the heritage of the KITSUNE CCB, capable of storing, processing and managing the large volumes of data. Equipped with a Raspberry Pi Compute Module 4, the VERTECS CCB has been designed to allow for high-speed communication with the camera payload and X-band transmitter. The inclusion of the Raspberry Pi CM4 should also encourage the use of more complex algorithms and onboard machine learning, thanks to its impressive processing power. Developed using open-source software and Commercial Off-The-Shelf components, the VERTECS CCB can be affordably fabricated and easily reused. The ability to meet the system requirements of the VERTECS project has been validated through tests performed in the table-sat configuration, as well as during the thermal vacuum test. The interfaces with each satellite subsystem have been confirmed, as well as the survivability of the COTS components. However, further work is still required to complete the mission software and to ensure that data integrity is preserved end-to-end through the payload pipeline. The VERTECS CCB will be available online as an open-source project, to enable future nanosatellite missions.

\acknowledgments 
The authors acknowledge the support, contributions and efforts of the entire VERTECS team for making this work possible. The authors also acknowledge the developers of the KITSUNE CCB: Abhas Maskey, Necmi Cihan Orger, Muhammad Hasif Bin Azami. The KITSUNE CCB served as a valuable source of reference during the design of the VERTECS CCB. Special thanks goes to the JAXA-SMASH program for enabling the VERTECS project.

\bibliography{report} 
\bibliographystyle{spiebib} 

\end{document}